# LIGHT CHEMICAL ELEMENTS IN STARS: MYSTERIES AND UNSOLVED PROBLEMS

L. S. Lyubimkov

*The first eight elements of the periodic table are discussed: H, He, Li, Be, B, C, N, and O. They are referred to as key elements, given their important role in stellar evolution. It is noteworthy that all of them were initially synthesized in the Big Bang. The primordial abundances of these elements calculated using the Standard Model of the Big Bang (SMBB) are presented in this review. The good agreement between the SMBB and observations of the primordial abundances of the isotopes of hydrogen and helium, D, $^3$He, and $^4$He, is noted, but there is a difference of ~0.5 dex for lithium (the isotope $^7$Li) between the SMBB and observations of old stars in the galactic halo that has not yet been explained. The abundances of light elements in stellar atmospheres depends on the initial rotation velocity, so the typical rotation velocities of young Main Sequence (MS) stars are examined. Since the data on the adundances of light elements in stars are very extensive, the main emphasis in this review is on several unsolved problems. The helium abundance He/H in early B-type MS stars shows an increment with age; in particular, for the most massive B stars with masses $M$ = 12– 19 $M_\odot$, He/H increases by more than a factor of two toward the end of the MS. Theoretical models of stars with rotation cannot explain such a large increase in He/H. For early B- and late O-types MS stars that are components of close binary systems, He/H undergoes a sharp jump in the middle of the MS stage that is a mystery for the theory. The anomalous abundance of helium (and lithium) in the atmospheres of chemically peculiar stars (types He-s, He-w, HgMn, Ap, and Am) is explained in terms of the diffusion of atoms in surface layers of the stars, but this hypothesis cannot yet explain all the features of the chemical composition of these stars. The abundances of lithium, beryllium, and boron in FGK-dwarfs manifest a trend with decreasing effective temperature $T_{eff}$ as well as a dip at $T_{eff}$ ~ 6600 K in the Hyades and other old clusters. These two effects are among the unsolved problems. In the case of lithium, there is special interest in FGK-giants and supergiants that are rich in lithium (they have log $\varepsilon(Li)$ > 2 ). Most of them cannot be explained in terms of the standard theory of stellar evolution, so nonstandard hypotheses are invoked: the recent synthesis of lithium in a star and the engulfment by a star of a giant planet with mass equal to that of Jupiter or greater. An analysis of the abundances of carbon, nitrogen, and oxygen in early B- and late O-type MS stars indicates that the C II, N II, and O II ions are overionized in their atmospheres. For early B-type MS stars, good agreement is found between observations of the N/O ratio and model calculations for rotating stars. A quantitative explanation of the well-known "nitrogen-oxygen" anticorrelation in FGK-giants and supergiants is found. It reflects the dependence of the anomalies in N and C on the initial rotation velocity $V_0$. The stellar rotation models which yield successful explanations for C, N. and O cannot, however, explain the observed increased helium enrichment in early B-type MS stars.*

## 1. Introduction

The first eight elements of the periodic table, H, He, Li, Be, B, C, N, and O, can have significantly different observed abundances during stellar evolution, beginning with the first, longest lasting stage when hydrogen burns in the core of a star; this is the Main Sequence (MS) stage. They are often referred to as key elements, given their key role in understanding the evolution of stars. These elements are also interesting in that their primordial synthesis took place in the Big Bang and that observational data on the primordial abundance of their isotopes is available which can be compared with cosmological models.

These elements are listed in Table 1, which, besides their atomic number, lists their most widespread isotope, ionization potential $E_{ion}$, and solar abundance log ε(El) [1]. All of the concentrations are given relative to hydrogen, the most widespread element in the observed universe. log ε(El) is given here on the standard logarithmic scale, where it is assumed that log ε(H) =12.00 for hydrogen.

The author has already published reviews on helium and lithium, in which data on the observed abundances of He and Li (relative to H) in stellar atmospheres are summarized, and their consistency with the predictions of the theory of stellar evolution is examined [2,3].

Published data on the observed abundances for the light elements from Table 1, as well as their theoretical interpretation, are very rich in the case of Li, C, N, and O, less numerous for He and Be, and rather sparse for B. On the whole, however, the material for discussion is so extensive that a detailed analysis of it would require writing a separate book. Thus, in this review, primary attention is devoted to just a few problems which are a mystery for the modern theory or, at least, still do not have a generally accepted explanation.

TABLE 1. A List of the Eight Lightest Elements and their Abundances in the Sun's Atmosphere (Asplund, et al. [1])

| Element | Atomic number | Principal isotope | $E_{ion}$, eV | logε(El) |
|---|---|---|---|---|
| H | 1 | $^1$H | 13.60 | 12.00 |
| He | 2 | $^4$He | 24.59 | 10.99* |
| Li | 3 | $^7$Li | 5.39 | 1.05 |
| Be | 4 | $^9$Be | 9.32 | 1.38 |
| B | 5 | $^{11}$B | 8.30 | 2.70 |
| C | 6 | $^{12}$C | 11.26 | 8.43 |
| N | 7 | $^{14}$N | 14.53 | 7.83 |
| O | 8 | $^{16}$O | 13.62 | 8.69 |

\* Helium lines are not observed in the sun's photospheric spectrum; the helium abundance given here corresponds to the average abundance in nearby young B-type stars [2].

We note that the observed abundances of the light elements discussed below were determined without the condition of LTE (local thermodynamic equilibrium), or at least where departures from LTE play a significant role.

## 2. Primordial abundance of the light elements

**2.1. Calculations based on the SMBB.** Mysteries show up at the very beginning, in discussions of the consequences of the Big Bang. It is remarkable that all eight of the elements were originally synthesized in the Big Bang. Table 2 lists current data [4] on the primordial abundances of these elements based on the Standard Model of the Big Bang (SMBB). In the third column of the table these data are given in the form of the abundance $\log\varepsilon(\text{El})$ reduced to a logarithmic scale.

It is interesting that, as opposed to earlier work, where the SMBB predicted the synthesis of only five of the first light elements (H, He, Li, Be, and B), the modern data include primordial nucleosynthesis of C, N, and O, as well. Although the usual yield of these elements is very small (the combined $CNO/H \sim 7\times10^{-16}$), even such a tiny abundance of these could play a decisive role in the evolution of the very first stars.

The highest yields in the Big Bang were of hydrogen, helium, and lithium. Observational data exist on the primordial abundances of the isotopes of these three elements: they are represented in the last column of Table 2. Here the relative abundance D/H of deuterium was determined from observations of intergalactic clouds of neutral hydrogen lying along the path of radiation from quasars with a large red shift, what is known as "damped Lyα systems" [5]. The primordial abundance of $^4$He was determined from observations of H II regions in old dwarf galaxies with low metallicities [6]. The primordial abundance of $^3$He was estimated from observations of H I regions in our galaxy [7]. Only the primordial abundance of lithium, more precisely of its more widespread isotope 7Li, was determined from stars – old FGK-dwarfs in the galactic halo [8,9].

TABLE 2. Primordial Abundances of Isotopes of the Eight Light Elements Calculated on the Basis of the SMBB [4]

| Quantity | Abundance | $\log\varepsilon(\text{El})$ | $\log\varepsilon(\text{El})$ observations |
|---|---|---|---|
| D/H | $2.59\times10^{-5}$ | 7.41 | 7.403±0.007 [5] |
| $^4$He/H | $8.23\times10^{-2}$ | 10.92 | 10.932±0.005 [6] |
| $^3$He/H | $1.04\times10^{-5}$ | 7.02 | 7.04±0.08 [7] |
| $^7$Li/H | $5.24\times10^{-10}$ | 2.72 | 2.2±0.1 [8,9] |
| $^6$Li/H | $1.23\times10^{-14}$ | -1.91 | |
| $^9$Be/H | $9.60\times10^{-19}$ | -6.02 | |
| $^{10}$B/H | $3.00\times10^{-21}$ | -8.52 | |
| $^{11}$B/H | $3.05\times10^{-16}$ | -3.52 | |
| $^{12}$C/H | $5.34\times10^{-16}$ | -3.27 | |
| $^{13}$C/H | $1.41\times10^{-16}$ | -3.85 | |
| $^{14}$C/H | $1.62\times10^{-21}$ | -8.79 | |
| $^{14}$N/H | $6.76\times10^{-17}$ | -4.17 | |
| $^{15}$N/H | $2.25\times10^{-20}$ | -7.65 | |
| $^{16}$O/H | $9.13\times10^{-20}$ | -7.04 | |
| CNO/H | $7.43\times10^{-16}$ | -3.13 | |

Table 2 shows that the observed abundances of the isotopes of hydrogen and helium, i.e., deuterium, $^4$He, and $^3$He, are in very good agreement with the SMBB calculations. The situation is utterly different for lithium: the observed abundance $\log\varepsilon(^7\text{Li}) = 2.2$ is smaller than the theoretically predicted value of 2.7 by 0.5. Thus, data on the primordial abundance of lithium derived from observations of old stars in the galactic halo require a separate analysis.

**2.2. Lithium in old stars in the galactic halo.** The first observations of lithium in the spectra of old FGK-dwarfs in the galactic halo yielded an interesting fact: it turned out that these stars, which have low metallicity [Fe/H] < –1, have a lithium abundance that is surprisingly constant. This constancy of $\log\varepsilon(\text{Li})$, for [Fe/H] ranging between -1 and -3, came to be known as the "lithium plateau" or the "Spite plateau," based on the name of the two French astronomers who first discovered this phenomenon 35 years ago [10]. The first image of the "lithium plateau" from their paper is reproduced in Fig. 1, where the abundances $\log\varepsilon(\text{Li})$ are plotted as a function of effective temperature $T_{eff}$. Here a value of $\log\varepsilon(\text{Li}) = 2.05 \pm 0.15$ was found for the "lithium plateau," while the current value is $\log\varepsilon(\text{Li})=2.2$, as noted above [8,9].

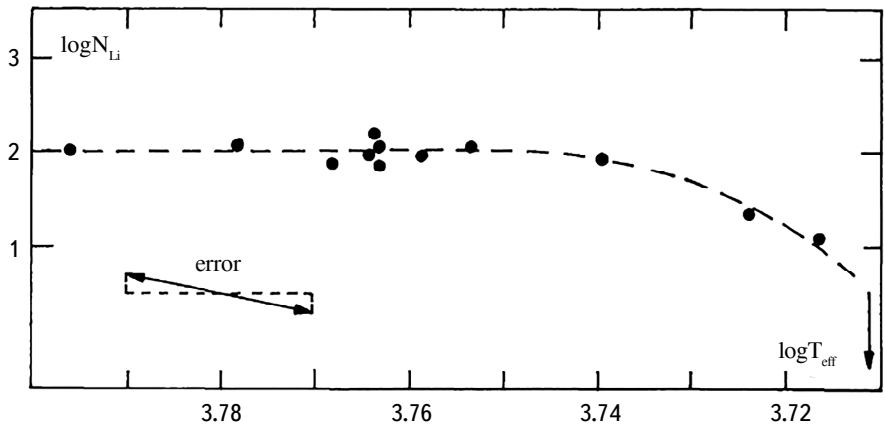

Fig. 1. The first representation of the "lithium plateau" in Spite and Spite [10].

It is significant that the value $\log\varepsilon(\text{Li}) = 2.2$ is smaller by 0.5 than the value of 2.7 predicted by the SMBB. This discrepancy has long been known, and all attempts to eliminate it by improving the SMBB have been in vain. In particular, Cyburt, et al. [11], have recently obtained primordial abundances of $^7\text{Li}/\text{H} = 4.648 \times 10^{-10}$ and $^6\text{Li}/\text{H} = 1.288 \times 10^{-14}$ using current values of the nuclear reaction rates; these values correspond to $\log\varepsilon(^7\text{Li}) = 2.67$ and $\log\varepsilon(^6\text{Li}) = -1.89$ on the standard scale. They are essentially the same as the data of Table 2. Thus, the discrepancy of -0.5 dex between theory and observations for the primordial abundance of $^7$Li remains.

As new data have been accumulated on the abundance of lithium in the atmospheres of old FGK-dwarfs in the galactic halo, the concept of a "lithium plateau" has undergone significant changes. The current distribution of $\log\varepsilon(\text{Li})$ with respect to $T_{eff}$ and [Fe/H] for these stars is shown in Fig. 2 ([3], based on data from [12]). The main change compared to Fig. 1 is that the "frozen" horizontal line at $\log\varepsilon(\text{Li}) \sim 2$ has begun to "thaw," as if spilling in droplets (the downward arrows indicate that only an upper limit could be estimated for the the Li abundance). Here about a dozen of the stars have the lithium abundance $\log\varepsilon(\text{Li}) \leq 1.5$.

It should be noted that, based on a large sample of stars, Spite, et al. [9], have shown recently that within a fairly limited range of metallicity −2.8 < [*Fe/H*] < −2.0, the stars, as before in Fig. 1, are grouped near a "lithium plateau" with $\log\varepsilon(Li) = 2.2$.

Despite repeated attempts to solve the problem of the lithium abundance in old stars of the galactic halo, as before two important questions remain as mysteries for the theory:

(1) why is there a discrepancy of 0.5 dex between the observed lithium abundance and the primordial abundance predicted by the SMBB?

(2) why do a significant fraction of the halo dwarfs have a lithium abundance significantly below the "lithium plateau?"

Until these questions are answered reliably, the lithium abundance problem for old stars cannot be regarded as solved.

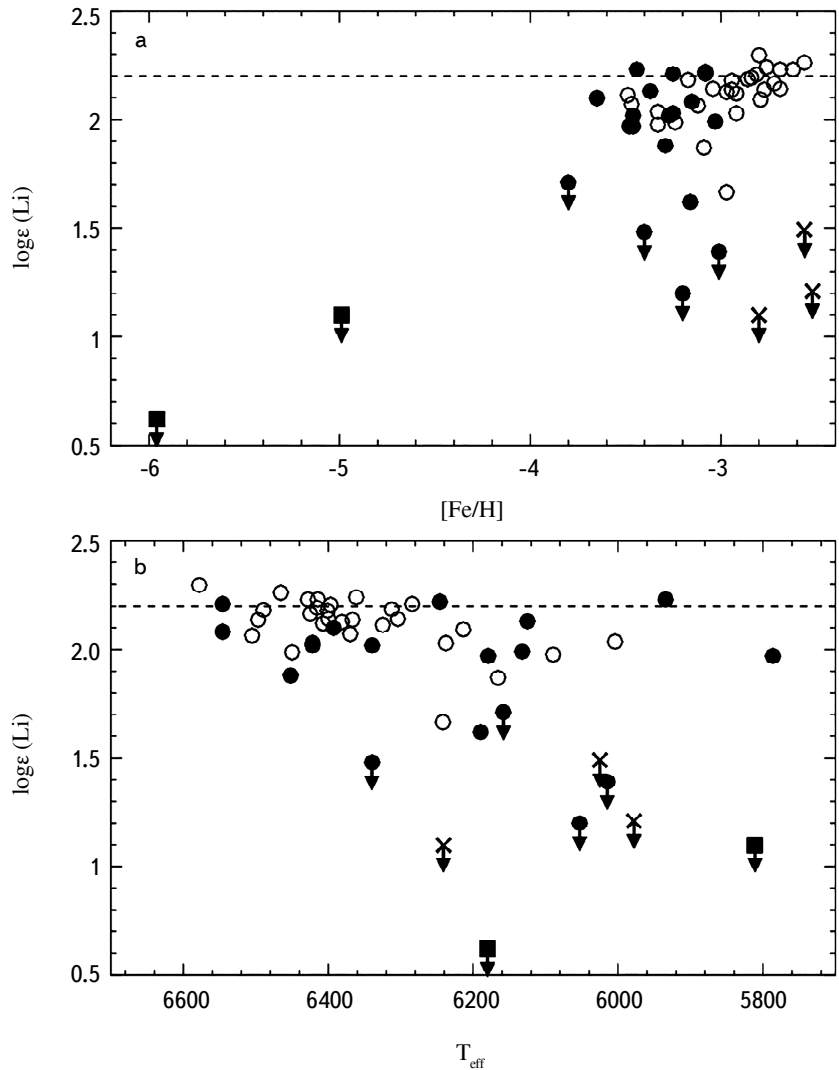

Fig. 2. Up-to-date distributions of the Li abundance for halo stars with respect to (a) the metallicity index [*Fe/H*] and (b) the effective temperature $T_{eff}$ [3]. The data points with arrows correspond to upper limits on $\log\varepsilon(Li)$. The dashed line corresponds a lithium abundance of $\log\varepsilon(Li) = 2.2$.

## 3. Rotation velocities of young stars

From the old stars in the galactic halo we proceed to the young stars in the thin disk. The evolution of these stars, including the evolution of the light elements in their atmospheres, depends both on a star's mass $M$ and its initial rotation velocity $V_0$. The importance of rotation became understandable during the transition from calculations with the traditional stellar models neglecting rotation to rotating star models [13]. Mixing occurs in a star because of rotation and it can lead to a change in the amounts of various chemical elements at its surface. Calculations showed that this effect is stronger then the mass M and rotational velocity $V_0$ are higher.

The effect of the mixing induced by rotation for stars with masses $M > 2\,M_\odot$ shows up by the MS stage. For these stars, reactions in the CNO-cycle in the core of a star play a dominant role and here the mixing owing to rotation shows up in two ways. On one hand, products of the CNO-cycle are transported from the interior of a star to the surface, so that the amounts of He and N in the atmosphere are enhanced, while the abundance of C is reduced. On the other hand, Li, Be, and B atoms are transported from the star's surface into deeper and hotter layers, where they are consumed in (p, $\alpha$) reactions, so that the abundances of Li, Be, and B in the star's atmosphere are reduced.

The stars discussed below have masses in the range from 1 to 40 $M_\odot$ (here $M_\odot$ is the sun's mass). Evolu-tionary calculations show that stars with masses $M \geq 20\,M_\odot$ in the initial MS (i.e., in the Zero Age Main Sequence, ZAMS) are O-type stars, while stars with $M \approx 4-19\,M_\odot$ on the ZAMS correspond to early B-type MS stars. Stars with $M = 2-3\,M_\odot$ on the ZAMS are late B- and A-stars, while objects with $M = 1-2\,M_\odot$ on the ZAMS are F- and G-dwarfs. Since the initial rotation velocity $V_0$ plays an important role in the following discussion, this raises the following question: what are the actual rotation velocities observed in these stars at the beginning of the MS phase?

It should be noted that calculations with rotating star models show that the rotation velocity should decrease during the MS stage. For example, the calculated rotation velocity toward the end of the MS can decrease by 10-20%, or even 100% [14]: the exact value depends on $V_0$ and $M$. However, a radial reduction in the rotation velocity occurs later, when a star leaves the MS and enters the cold giant or supergiant stage. Thus, as a first approximation we can assume that the changes in the rotation velocity along the MS are negligible; then the question stated above can be formulated in a more general way: what are the rotation velocities observed in MS stars?

The projection of the rotation velocity at the equator $V$ along the line of sight, $V\sin i$, is determined from observations. Note that for the transition from $V\sin i$ to $V$, it is necessary to take into account the fact that the average value of the random quantity $\sin i$ equals $\pi/4$. [15]. Current data, especially for hot class O and B stars, can be used for a substantial reexamination of the representations of $V\sin i$ and $V$ of stars in the MS stage available 30 years ago (for example, in Allen's handbook [15]). In brief, they reduce to the following.

For stars in classes O and B (as opposed to previous representations), the fraction of stars with relatively slow rotation is large [16-18]. In fact, most early B- and O-type stars (more precisely about 70-80%) had low rotation velocities ~0-150 km/s at the beginning of evolution on the MS. Here a substantial fraction of them fell within a still narrower interval, 0-50 km/s. The number of these stars with relatively high rotation velocities from 150 to 300-400 km/s was small: ~20% for early B stars with masses $M \approx 4-19\,M_\odot$ and ~30% for late O-type stars with $M = 20-40\,M_\odot$. The later B-type MS stars and A-stars with masses $M$ from 4 to 2.5 $M_\odot$ have a bimodal distribution with peaks at 50 and 260 km/s with a full range of values of $V$ from 0 to 300-400 km/s [16]. The late A- and early F-stars with masses $M$ from 2.0 to 1.6 $M_\odot$ had a unimodal distribution of the velocity $V$ with a peak at ~150 km/s [19].

With further reduction in $M$, a rapid reduction in $V$ is observed. A sharp drop in V from ~150 to 10 km/s has been detected for F-stars with $M \approx 1.4 M_\odot$ (dwarfs in subclass F4). We note that this phenomenon is associated with well-known "Li and Be dip" at $T_{eff}$~6600 K in old clusters, e.g., in the Hyades (see section 5.1). For less massive stars with $M \leq 1 M_\odot$ on the MS, low rotation velocities <10 km/s are typical. Recall that the sun, a G2V dwarf, has a rotational velocity at the equator of $V$ = 2 km/s [15].

As will be seen in the following, these features of the rotation velocity distribution of young stars are important for interpreting the observed abundances of light elements, both for stars in the MS stage and for those in the subsequent AFG-supergiant stage.

## 4. The "helium/hydrogen" ratio

Helium, element number two in the periodic table is the second element after hydrogen in terms of its dispersion throughout the observed universe (in stars and gaseous nebulae). Two important quantities regarding helium have to be established. First, the primordial "helium/hydrogen" ratio (in terms of the number of atoms) is He/Hp = 0.082 (Table 2) and second, the current initial abundance of helium for young B-type MS stars in the vicinity of the sun is He/H = 0.098±0.003 [2]. The interstellar medium has been enriched in helium by roughly 20% over the lifetime of the galaxy and models of its chemical evolution show that this has been caused mainly by massive type II supernova explosions.

Data on the abundance of the dominant helium isotope $^4$He are discussed in the following. The contribution of the other isotope $^3$He to the abundance of helium is very small; e.g., for the sun $^3\text{He}/^4\text{He} = 1.7 \times 10^{-4}$ [1].

**4.1. Helium enrichment of the atmospheres of early B-type MS stars.** Because of its high ionization potential (Table 1), the lines of neutral helium are observed only in the spectra of hot stars from O to early A. Especially complete data on the abundance of helium have been obtained for early B-type MS stars, where the He I lines are particularly strong.

It is known that in the MS stage the main source of energy for stars with masses $M$ >2 $M_\odot$ is the CNO-cycle. In this state, hydrogen burns in the core of a star and is converted into helium; here the He/H ratio in the interior of a star increases strongly (toward the end of the MS stage, the hydrogen in the core of a star is almost completely converted to helium). It turned out that He/H can simultaneously increase in the atmosphere of early B-type MS stars. Evidence of this was obtained by the author 40 years ago [20,21], but then it was entirely unexpected for the theory. Later it became clear that the reason for the observed increase in He/H in the main sequence may be mixing induced by rotation of a star that leads to transport of products of the CNO-cycle (including helium) from the interior to the surface.

Fairly complete data on helium enrichment of the atmospheres of B-type MS stars were obtained by Lyubimkov, Rostopchin, and Lambert [22] in a non-LTE analysis of the helium abundance for 102 early B-type MS stars (two of which were chemically peculiar "He-weak" stars). The masses of the 100 stars subjected to further analysis ranged from 4 to 19 $M_\odot$, and their observed rotation velocities $V\sin i$, found [22] using the same six He I lines that were used to determine the helium abundance, ranged from 0 to 280 km/s. These data showed that during the MS stage, He/H increases with the age of the stars and this effect tends to increase with the mass $M$ and rotation velocity.

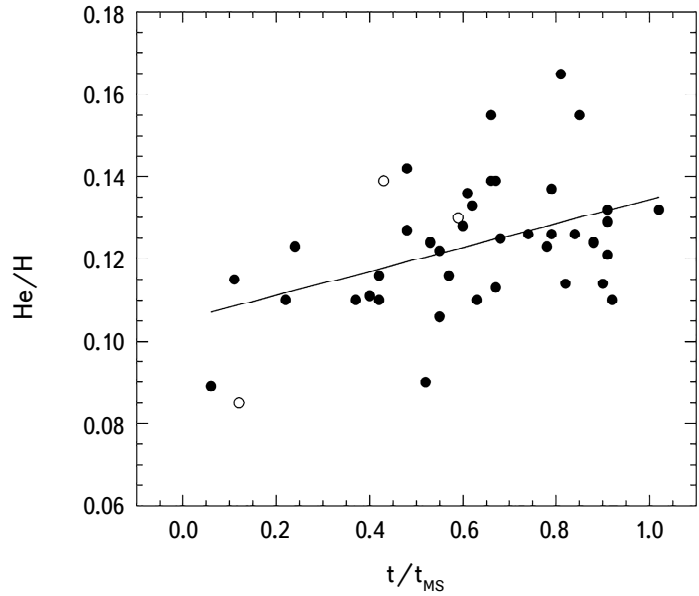

Fig. 3. He/H as a function of relative age for B-type MS stars with masses $M = 4 - 7 M_\odot$.

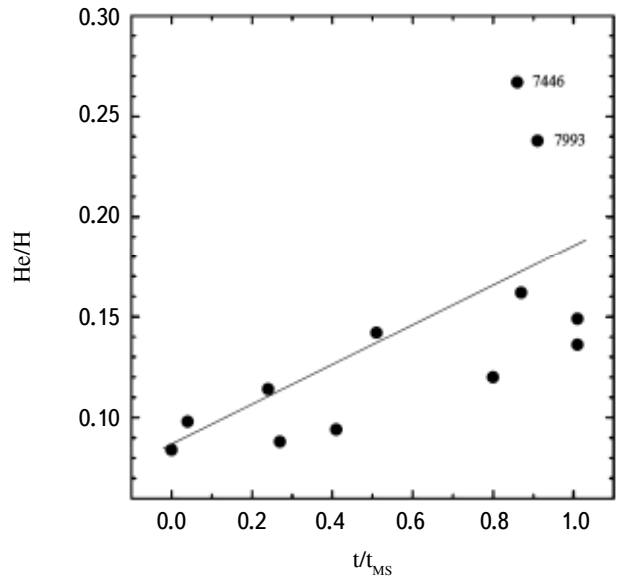

Fig. 4. He/H as a function of relative age for B-type MS stars with masses $M = 12 - 19 M_\odot$.

Figures 3 and 4 (from [22]) show the dependence of He/H on the relative age $t/t_{MS}$ (here $t$ is the age and $t_{MS}$ is the lifetime of a star of given mass on the MS) for two groups of stars with different masses $M$: stars with $M = 4-7 M_\odot$ (Fig. 3) and more massive stars with $M = 12-19 M_\odot$ (Fig. 4). For the first group, the increase in He/H over the MS stage averages 28%, and for the second group an increase in He/H by more than a factor of two toward the end of the MS was observed. Two of the giants, HR 7446 and 7993, had especially high helium abundances, He/H = 0.27 and 0.24, apparently because of their high rotation velocities ($V\sin i = 270$ and 224 km/s, respectively).

As a whole, Figs. 3 and 4 seem to agree with the computational models for rotating stars. The theory predicts that the enrichment of the atmosphere in helium during the MS stage is greater for higher masses and rotation velocities of the stars. In terms of quantitative estimates, however, there is no agreement with the theory.

For example, current calculations [14] predict that for a model with $M = 15 M_\odot$, even with an initial rotation velocity $V_0 = 500$ km/s (0.9 times the critical velocity), the increase in He/H in the atmosphere will only be by 28%, which is inconsistent with the observed increase by a factor of two in He/H for stars with $M = 12-19 M_\odot$ (Fig. 4).

In the meantime, as shown in section 3, most (~80%) of the early B-type MS stars with masses $M = 12-19 M_\odot$ have relatively low rotation velocities of 0-150 km/s. It is important that the data of [22] are in full agreement with this conclusion; in fact, of the 100 stars studied there, 85 (i.e., 85%) have $V\sin i \leq 150$ km/s and 15 have $V\sin i > 150$ km/s, while only 8 stars have $V\sin i > 200$ km/s. But, according to calculations [14], only for rotation velocities of 400-500 km/s is a noticeable increase in He/H in the atmosphere of a star possible.

Thus, there are some serious disagreements between the theory and the observations. The observational data on the increased helium abundance in the atmospheres of early B-type MS stars suggest that the theoretical models seriously underestimate the abundance of helium (the main product of the CNO-cycle) transported to the surface of a star by mixing. In other words, it can be assumed that the entrainment of helium starts at much lower rotation velocities than found in modern theoretical models.

*Therefore, explaining the observed helium enrichment of the atmosphere of early B-type MS stars is still an unsolved problem for the theory.*

**4.2. Helium in close binary systems.** An utterly unique behavior of He/H is observed in hot stars which are components of close binary systems. Evidence of this is provided by Fig. 5, which shows the dependence of He/H on the relative age $t/t_{MS}$ for the components of five binary B-type MS stars studied at the Crimean Astrophysical Observatory (CrAO) [23] (solid circles). Here the basic parameters $T_{eff}$ and $\log g$, as well as the helium abundance He/H, were determined separately for each component.

Figure 5 also shows the values of He/H for two binary O-type stars [24,25] (hollow circles). All these systems have orbital periods ranging from 2 to 14 days, and they have not yet reached the phase in which mass is exchanged between the components.

Figure 5 shows that, as opposed to the gradual, monotonic increase in He/H observed in isolated hot stars (Figs. 3 and 4), the low (original) helium abundance is retained in the components of close binary systems throughout the first half of their evolution on the Main Sequence ($t/t_{MS} < 0.5$). Then, over the short interval of $t/t_{MS}$ from 0.5 to 0.7, there is a sharp increase in He/H by roughly a factor of two, after which He/H remains at this higher level until the end of the MS stage.

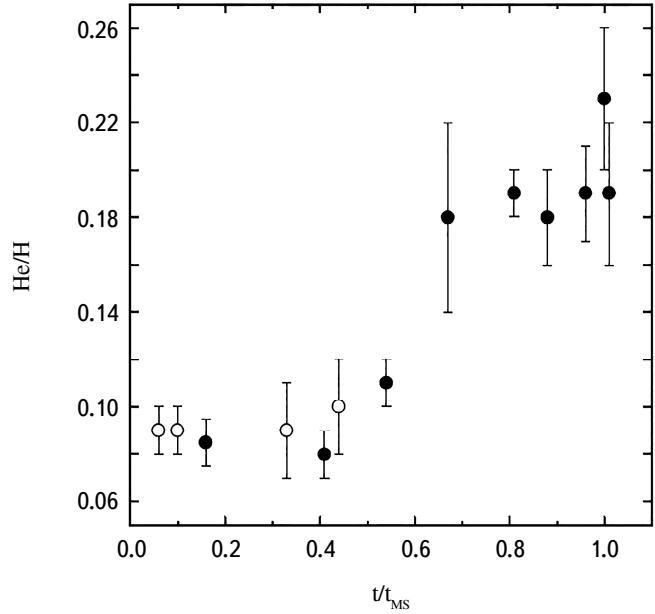

Fig. 5. He/H as a function of relative age for early B- and O-type MS stars that are components of close binary systems.

*For the modern theory this kind of sudden rise in He/H in the atmosphere in the middle of the MS stage is a mystery.*

Calculations of mixing in the MS are needed both for isolated stars and for the components of close binary systems where, besides rotation, the tidal interactions of the components must be taken into account.

**4.3. Helium in the atmospheres of chemically peculiar stars.** Chemically peculiar stars (CP-stars) of spectral types A, B, and F, for which the abundances of elements are observed to differ significantly from stars with the normal (solar) chemical composition, have been of special interest for several decades. The helium abundance in them also exhibits anomalies. Table 3 lists five types of CP stars (from [2]). Here the characteristic ranges of $T_{eff}$ and He/H from the literature are listed for each type. Evidently, all these stars are in the MS stage.

Two types of CP stars with different helium anomalies are observed in class B: "He-strong" (He-s or He-r=He-rich) stars with enhanced helium lines and "He-weak" (He-w) stars with attenuated helium lines. A high helium abundance He H $\approx 0.3/-10$ is observed in the first type and a low abundance He H $\approx 0.005 - 0.05$ in the second. It is interesting that for $T_{eff}$ between 17000 and 25000 K, three different types of B-type MS stars are observed in the Main Sequence stage: He-s, He-w, and normal B-type MS stars. This phenomenon is one of the unsolved problems for CP-type stars.

TABLE 3. Typical Effective Temperatures $T_{eff}$ and Helium Abundances He/H for Five Types of Chemically Peculiar Stars

| Type of stars | Range of $T_{eff}$ (approximately) | Range of He/H |
|---|---|---|
| He-strong | 17000-32000 | 0.3-10 |
| He-weak | 13000-25000 | 0.05-0.005 |
| HgMn | 10000-14000 | 0.05-0.005 |
| Am | 7300-10000 | 0.03-0.006 |
| Magnetic Ap | 7500-11000 | < 0.05 |

It is evident that an explanation of the anomalous abundance of helium in CP-stars should be sought together with an explanation of other features of the chemical composition of these objects, including the following:

(1) For magnetic Ap-stars and type Am and HgMn stars there is a well known trend in the behavior of the chemical anomalies: the excesses of elements increase on the average with increasing atomic number $Z$ [26]. While these excesses are usually small or even have a "minus" sign for the light elements, with increasing $Z$ the excesses increase up to 6-7 dex for the heaviest elements [27].

(2) For CP-stars with significant magnetic fields (Ap, He-s, and some He-w stars), a nonuniform distribution of the elements over the star's surface is typical. Spots of some elements, including helium and lithium, lie in the regions of the magnetic poles, while spots of the other elements are either concentrated at the magnetic equator or have no systematic positions.

(3) The characteristic feature of CP-stars is stratification of the elements, i.e., their abundances have a strong dependence on depth. Here different elements can manifest fundamentally different behavior: some have large excesses in higher layers of the atmosphere, while others have a deficit in these layers with normal or enhanced abundances in deeper layers (this depends on the element, as well as on the particular type of CP-star).

A diffusion hypothesis has been proposed as the basic explanation for these features of the composition of CP-stars. This means that these features are not associated with thermonuclear processes in the interior of a star, but are a result of the diffusion of atoms in its surface layers under the influence of two oppositely directed forces—gravitation and radiation pressure. In magnetic CP-stars, the magnetic field also contributes.

Detailed calculations of diffusion for any specific CP-star which might provide a quantitative explanation for all the observed phenomena, including the general trend toward excesses of all the observed elements with increasing $Z$, a map of the distribution of spots of different elements over the surface of a star, and the distributions of different elements with depth, are a difficult problem. It has not yet been solved completely for any individual CP-star.

## 5. Lithium, beryllium, and boron

Lithium, beryllium, and boron are a unified group of elements from the standpoint of evolution, since they burn up in stars in the same process, (p, α) reactions. But this process begins at different temperatures T~2.5, 3.5, and $5.0\times10^{6}$ K for Li, Be, and B, respectively. Thus, a change in the observed abundance of Be and, especially, B will require deeper mixing that reaches the hotter layers of a star than for lithium.

This leads to the conclusion that beryllium and boron are much less sensitive indicators of evolution than lithium. This result is confirmed by observations, as well as theory. For example, model calculations [28] for a star with mass $12\,M_\odot$ and an initial rotation velocity of 100 km/s showed that by the end of the MS the change in the lithium abundance in the atmosphere is -3.0 dex, for beryllium, -1.5 dex, and for boron, only -0.5 dex.

There is a very rich literature on lithium, many fewer publications on beryllium, and comparatively few on boron (the latter require UV observations from space). These data yield some interesting conclusions.

**5.1. Features of the abundances of lithium, beryllium, and boron in the atmospheres of FGK-dwarfs in the galactic disk.** The initial lithium abundance for stars in the thin disk in the vicinity of the sun is $\log\varepsilon(Li)=3.2$ [3]. This is an order of magnitude greater than the "relict" value $\log\varepsilon(Li)=3.2$ (the "lithium plateau") found for old dwarfs in the galactic halo (see section 2.2). The following question arises: how could this additional lithium show up in the disk?

Modern models of the chemical evolution of the universe give an answer to this question [3]. The interstellar medium has been enriched in lithium (isotope $^7$Li) through outflow or ejection of matter from stars of the following types: low-mass red giants (up to 40%), AGB (asymptotic giants branch), and novae. About 20% of the isotope $^7$Li (and 100% of the $^6$Li) was provided by galactic cosmic rays which produced spallation reactions on the heavier and much more numerous C, N, and O nuclei in the interstellar medium.

During evolution on the MS stages, the abundance of lithium in the atmospheres of dwarfs of spectral types F, G, and K changes substantially relative to the initial value, $\log\varepsilon(Li)=3.2$. This process depends on the effective temperature $T_{eff}$ (i.e., in fact on the mass $M$) and on the age, specifically: the Li abundance decreases with age, and more rapidly so if $T_{eff}$ or $M$ are lower. In particular, over the sun's lifetime t = 4.5 billion years, according to current estimates [1,29,30] listed in Table 4, the abundance of lithium in its atmosphere has fallen by roughly a factor of 140 relative to the initial value $\log\varepsilon(Li)=3.2$.

TABLE 4. Current estimates of the Lithium Abundance in the Sun's Photosphere Based on 3-D Hydrodynamic Models of the Solar Atmosphere

| $\log\varepsilon(Li)$ | Reference |
|---|---|
| 1.05±0.10 | Asplund, et al. [1] |
| 1.03±0.03 | Caffau, et al. [29] |
| 1.07±0.02 | Monroe, et al. [30] |

The dependence of $\log\varepsilon(\text{Li}) = 3.2$ on $T_{eff}$ shows up especially well in studies of stars in a single cluster, since their ages are approximately the same. Besides the trend in the Li abundance with $T_{eff}$ for FGK-dwarfs of old clusters, in particular for the Hyades cluster (age ~700 million years), yet another puzzling phenomenon has been discovered: a deep drop ("Li dip") in the distribution of the Li abundance near $T_{eff} \approx 6600$ [31].

Figure 6 [26] shows the distribution of the Li abundance for FGK-dwarfs in the Hyades, along with (for comparison) the distribution for dwarfs in the younger Pleiades cluster (age ~100 million years). It can be seen that the trend in the Li abundance with decreasing $T_{eff}$ is less distinct in the Pleiades and the lithium dip is entirely absent.

Later, an analogous, but not so deep, drop for beryllium ("Be dip") was discovered for stars in the Hyades [32]. It is interesting that in the younger, open clusters of the Pleiades and a Per (*t*~100 million years), no Be dip was found, while in the Coma Ber = Mel 111 cluster (t~500 million years), a Be dip comparable in depth to the dip in the Hyades exists [33]. This implies that the Be dip, like the Li dip, shows up in MS stars with masses $M = 1-2 M_\odot$ for ages ranging from 100 to 500 million years.

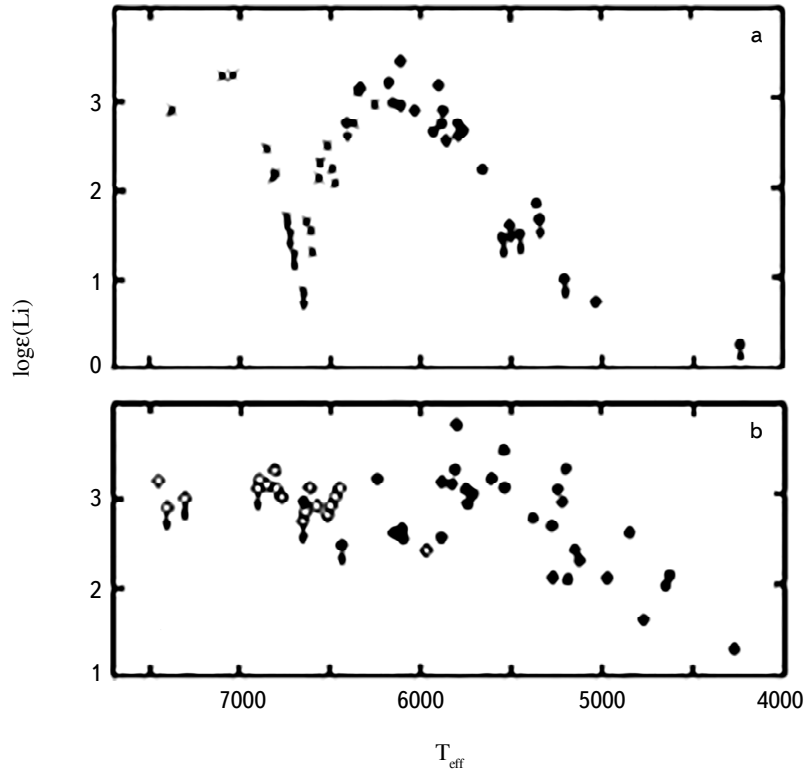

Fig. 6. Distribution of the lithium abundance with respect to effective temperature for FGK-dwarfs in the Hyades (a) and Pleiades (b) clusters [26].

Boesgaard, et al. [34], have recently obtained new data for stars in the Hyades which showed that, besides a lithium and beryllium dip, there is also a boron dip ("B dip"). Here the boron abundance for several stars was determined from the B I 2496.8 Å UV line (observed with the HST). The depths of the dip in these three cases were different: more than 2.0 dex for Li, about 1.0 dex for Be, and 0.4 dex for B. This difference is entirely to be expected, since, as noted above, of these three elements lithium is the most sensitive indicator of evolution, while beryllium and, especially, boron are less sensitive.

The dip in the distributions of Li, Be, and B near $T_{eff}$ = 6600 K for stars in the Hyades may be related to the fact (noted above in section 3) that a sharp drop in the rotation velocity $V$ from ~150 to 10 km/s is observed at this value of $T_{eff}$ in F-dwarfs in the Hyades [34].

The problems discussed above might be regarded as solved if some model calculations could reproduce the observed distributions of the abundances of Li, Be, and B with respect to $T_{eff}$, e.g., for the stars in the Hyades (as in Fig. 6). No such calculations, however, have yet been done.

*Therefore, the reduction in the abundances of lithium and beryllium in the atmospheres of FGK-dwarfs (including the sun, where Li is reduced by a factor of 140), as well as the trend with $T_{eff}$ and the dip in the abundances of Li, Be, and B in stars in old clusters such as the Hyades, still have no theoretical solution.*

**5.2. Lithium in the atmospheres of cold giants and supergiants.** An enormous number of papers are devoted to lithium. This rich material makes it possible, as opposed to the cases of beryllium and boron, to examine data on the the Li abundance for cold MS dwarfs, as well as for stars which are further evolved, such as the stages of FGK-giants and supergiants.

This stage is known to be accompanied by deep convective mixing (DCM), which leads to significant variations in the observed abundances of several light elements (recall that these kinds of changes can begin even in the MS stage if there was mixing caused by rotation). In particular, DCM causes an enhancement in the nitrogen abundance and a reduction in the oxygen abundance, while the abundance of lithium in an atmosphere may decrease to an undetectably low level. Here, as noted in [3], the changes in the Li abundance during the DCM phase begin earlier than those in the abundances of C and N.

The anticorrelation between the abundances of nitrogen and oxygen in FGK-giants and supergiants has long been known; it is discussed in section 6.3. As for lithium, it does not show up at all in the spectra of most of these stars. Thus, there is little lithium in the atmospheres of these stars or it has burnt up completely. Current models of rotating stars explain this fact completely.

These remarks can be illustrated by data from Refs. 35 and 36. The first reported the lithium abundance for 55 FGK-supergiants and giants, while the second raised the sample size to 146.

Figure 7 shows the lithium abundances $\log(\varepsilon Li)$ from these two papers as functions of mass $M$. The open symbols show the data [35] and the solid symbols, the data from [36]. The triangles represent an upper bound for $\log\varepsilon(Li)$ (that is, the lithium line is not observed in the spectra of these stars).

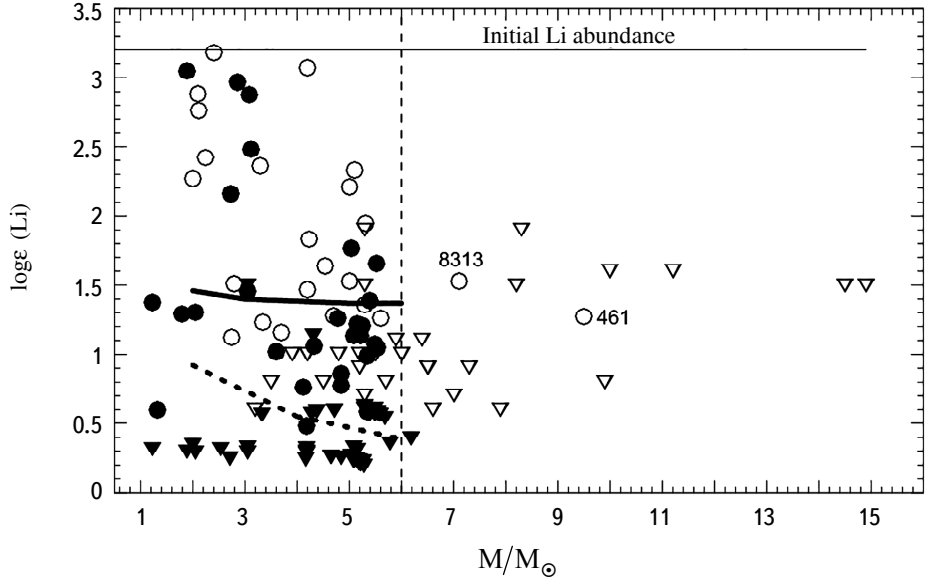

Fig. 7. Lithium abundance as a function of mass M for FGK-supergiants and giants. The open symbols correspond to 55 stars from [35] and the solid symbols, to an additional 91 stars from [36]. The triangles are an upper bound for $\log\varepsilon(\text{Li})$. The smooth and dashed curves are calculations for $V_0 = 0$ and 50 km/s, respectively.

These empirical data, as well as theoretical results, show that, in terms of the lithium abundance, FGK-supergiants and giants separate into two groups with masses $M \leq 6 M_\odot$ and $M > 6 M_\odot$.

For stars with $M \leq 6 M_\odot$, there is a large spread in the Li abundances from an initial value of $\log\varepsilon(\text{Li}) = 3.2$ to unobservably low values $\log\varepsilon(\text{Li}) < 1$. This large spread is related to the very high sensitivity of the atmospheric the Li abundance to the initial rotation velocity $V_0$. Changes can begin toward the end of the MS stage, even for low values of $V_0 \sim 50$ km/s. Calculations show that for $V_0 = 100$ km/s, the Li abundance falls by 3-4 dex at the end of the MS, i.e., it becomes unobservable (here the abundances of C and N in the atmosphere essentially are unchanged). The changes in the Li abundance during the FGK-giant/supergiant stage toward the end of the DCM phase are fairly large, even for $V_0 = 0$ and 50 km/s (see the smooth and dashed curves, respectively, in Fig. 7).

For stars with $M > 6 M_\odot$ the situation is simpler: here only low Li abundances are observed (Fig. 7); in most cases the Li line is entirely absent in the spectrum. Recall that most (~80%) of these comparatively massive stars have a rotation velocity below 150 km/s in the MS stage (see section 2). According to stellar model calculations, toward the end of the MS with $V_0 \approx 50 - 100$ km/s, almost all the lithium should vanish from the atmospheres of these stars. Furthermore, even for $V_0 = 0$ km/s, shortly after emergence from the MS stage there is sharp drop in the Li abundance [35]. Thus, in all stars with $M > 6 M_\odot$ practically all the lithium should burn up, even before their entry into the red giant/supergiant phase. Figure 7 shows that two stars in this group, the supergiants HR 461 (K0 Ia) and HR 8313 (G5 Ib) for which lithium has been detected at a level of $\log\varepsilon(\text{Li}) \sim 1.5$, are in conflict with this theory.

Thus, modern theoretical models with rotation fully explain the absence of lithium in the atmospheres of most FGK-giants and supergiants. They can explain the lithium abundance of $\log\varepsilon(Li) \approx 1-2$ for stars with masses $M \leq 6 M_\odot$. In the same group with $M \leq 6 M_\odot$, however, stars are observed which are rich in lithium with $\log\varepsilon(Li) \geq 2$; in most cases they present a mystery to the theory. In the group $M > 6 M_\odot$, as noted above, the lithium detected in the cold supergiants HR 461 and HR 8313 conflicts with the theory.

**5.3. The mystery of giants that are rich and super-rich in lithium.** Giants and supergiants that are rich in lithium (with $\log\varepsilon(Li) \geq 2$) are of heightened interest since most of them cannot be explained in terms of the standard theory of stellar evolution. These objects form a very small part of all FGK-giants and supergiants— from 1 to 3% according to various data. The fact that there are few of these stars may indicate either that this phase of evolution is very short or that the scenario of their origin is unusual.

Following the tendency in the literature, these stars can be divided into two subtypes: simply rich in lithium ("Li-rich giants") and "super Li-rich giants." The fundamental difference in the Li abundances in these two groups is that loge(Li) for the Li-rich giants does not exceed the initial value $\log\varepsilon(Li) = 3.2 \pm 0.1$, while for the super-rich Li giants $\log\varepsilon(Li) = 3.5 - 4.3$ which is substantially higher. It has been found that the giants and supergiants of both types have masses $M > 6 M_\odot$. This conclusion was based on an analysis [35] of all the available published data.

Models of rotating stars show that some fraction of the Li-rich supergiants can be successors to B-type MS stars with low initial rotation velocities of 0-50 km/s (under the condition that these stars have not passed through a DCM phase after completing the MS stage). The remaining Li-rich giants and all the super Li-rich giants are completely unexplainable in terms of the standard theory of stellar evolution. It should be noted that many of these stars have already passed through the DCM phase, as indicated by the low carbon isotope ratio $^{12}C/^{13}C$ in their atmospheres [3]; in this case, all the lithium in their atmospheres should have been burnt up.

There are two competing hypotheses for explaining stars that are rich in lithium. The first is recent synthesis of lithium after a DCM by the Cameron-Fowler mechanism [37]. It includes the reactions $^{3}He + \alpha \rightarrow {}^{7}Be + \gamma$ and $^{7}Be + e^{-} \rightarrow {}^{7}Li + \nu_e$; i.e., $^7$Li nuclei are synthesized from $^3$He via $^7$Be. In the Cameron-Fowler mechanism, convection should play an important role, since it facilitates rapid transfer of $^7$Be into colder layers of the atmosphere. As opposed to AGB stars with masses $M \approx 4 - 6 M_\odot$, where ordinary convection is sufficient for entrainment of $^7$Be into the upper layers, in the case of giants with $M \approx 1 - 2 M_\odot$ belonging to the red giant branch (RGB), extra mixing is required.

The second hypothesis is capture of a giant planet with mass equal to that of Jupiter or more by the star. This hypothesis has been discussed more actively in recent years because the rapidly increasing number of newly discovered exoplanets confirms that planetary systems around cold giants are a fairly widespread phenomenon. In addition, calculations show that migration of planets takes place in planetary systems as they form, so that a planet can be captured by the star.

It is interesting that this capture can (1) actuate the above mentioned Cameron-Fowler mechanism and (2) substantially increase the star's rotation velocity (anomalously high rotation velocities up to ~100 km/s, which are utterly atypical of FGK-giants, are actually observed in some Li-rich giants).

The hypothesis according to which a giant-planet is captured by a star is under active development. In particular, the consequences of the capture on a red giant of a planet with a mass up to 15 times that of Jupiter have recently been calculated [38]. It turns out that the lithium abundance on the star's surface can increase up to $\log\varepsilon(\text{Li}) \approx 2.2$ (neglecting the "extra mixing" effect). Since this is clearly not enough to explain the lithium abundance in most Li-rich giants, the Cameron-Fowler mechanism has to be taken into account along with extra mixing.

## 6. Carbon, nitrogen, and oxygen

These three elements participate in the CNO-cycle, which is the main source of energy in stars with masses $M > 2\,M_\odot$ in the MS stage. Changes in the abundances of C, N, and O in stars in different stages of evolution have been of great research interest for more than a decade. Lines of these three elements are observed over a wide range of spectral types from O to M. In particular, lines of C II, N II, and O II can be seen in the spectra of hot stars. Their intensity is highest in the spectra of early B-type MS stars.

**6.1. C, N, and O in the atmospheres of early B- and late O-type stars.** For a long time, different authors obtained reduced abundances of C, N. and O (relative to the sun) for early B-type MS stars based on lines of C II, N II, and O II. This discrepancy was gradually almost eliminated thanks to (1) more accurate estimates of the abundances of C, N, and O for B-type MS stars and (2) refined estimates of the abundances of C, N, and O for the sun based on nonstationary 3D hydrostatic models of the solar atmosphere. Nevertheless, a significant difference remained for carbon.

Studies in recent years have shown that young stars in the sun's vicinity have the same metallicity as the sun on the average. This has been shown for the abundances of N, O, Mg, Si, Fe, Cr, and Ti. Thus, the carbon deficit in early B-type MS stars could hardly be regarded as real. Most likely, it can be assumed that there are some kinds of defects in the calculations for C II lines.

Calculations of the C II, N II, and O II lines for early B- and O-type stars are based on standard plane-parallel models of stellar atmospheres. There is some doubt about their adequacy since they (1) cannot explain the observed x-ray emission from these stars and (2) yield a highly reduced UV flux in the region $\lambda < 912$Å (the H I continuum) and, especially, for $\lambda < 504$Å (the He I continuum). But this is the UV radiation that controls photoionization of the C II, N II, and O II ions. This is confirmed by UV observations of two early B-type stars, b CMa (B1 II-III) and e CMa (B2 II), with the EUVE satellite. The observed flux was greater than the theoretical value by two orders of magnitude for $\lambda < 504$Å and by several times for $\lambda < 912$Å.

It has been reported [39] that there appears to be superionization of C II, N II, and O II ions in the atmospheres of early B- and late O-type stars that is not taken into account in conventional calculations. It becomes significant at temperatures $T_{eff} > 18500$ K in the case of C II lines and $T_{eff} > 26000$ K for N II and O II lines. Neglecting this effect in such hot stars may reduce the abundances of C, N, and O by 0.2 dex. If we only consider B-type stars with temperatures $T_{eff}$ < 18500 K for determining the carbon abundance, then the above mentioned carbon deficit relative to the sun vanishes.

In some current papers the superionization problem in B-type MS stars is solved implicitly, with this phenomenon taken into account in determining the basis parameters of these stars: the effective temperature $T_{eff}$ and acceleration of gravity $\log g$. As opposed to the traditional method for determining $T_{eff}$ and $\log g$, which is based on using the photometric indices and Balmer lines, these papers are based exclusively on examining the ionization balance for the lines of several light elements, in particular C II-III, O I-II, and Ne I-II [40]. This means that for each star, the values of $T_{eff}$ and $\log g$ are chosen so that the abundance of a given element found for two neighboring ionization states should coincide. The parameters $T_{eff}$ and $\log g$ determined in this way were systematically high: $T_{eff}$ by up to 2300 K and $\log g$ by up to 0.5 dex. Because of this increase in $T_{eff}$ and $\log g$, the degree of ionization in the calculations is raised and the superionization problem is thereby eliminated. But another problem shows up: a systematic discrepancy between the values of $T_{eff}$ and $\log g$ found by fundamentally different methods.

In order to solve the superionization problem for hot stars to study the abundances of C, N, and O in their atmospheres, it is necessary to shift to more realistic models of their atmospheres: this assumes a transition from plane-parallel models to spherical models, as well as including stellar wind and the magnetic field that has been observed in a number of O- and B-type stars. In this regard, it is interesting to note that a magnetic field of ~100 G has recently been discovered [41] in the two B-giants with a strong UV excess mentioned above, b CMa and e CMa.

**6.2. The N/O ratio in the atmospheres of early B-type stars.** The previous section implies the following: in the case of early B-type MS stars, of the three ratios N/C, C/O, and N/O, which are regarded as indicators of stellar evolution, only N/O is reliable, since it was insensitive to superionization of N II and O II ions. On the other hand, N/C and C/O, which include carbon, may contain systematic errors owing to neglected superionization of C II ions in stars with effective temperatures $T_{eff}$>18500 K.

[N/O], the ratio N/O normalized to the initial value (on a logarithmic scale), has been determined for 46 early MS stars [42]. Figure 8 shows the resulting dependence of [N/O] on mass M for those stars which are at the end of the MS stage (with relative ages $t/t_{MS}$ = 0.70-1.02). Here the smooth curves show model calculations [14] for three values of the initial angular rotation velocity W relative to the critical velocity $\Omega_{crit}$. The relative angular velocity $\Omega/\Omega_{crit}$ and the corresponding linear velocity $V_0$ (this is an average, since $V_0$ depends on $M$) are indicated next to each curve.

Figure 8 shows that most stars at the end of the MS have relatively low $[N/O] \leq 0.3$ and correspond theoretically to a model with $\Omega/\Omega_{crit} = 0$ and 0.3 (or $V_0$ from 0 to 130 km/s). For the four stars (18%) with the highest values of [N/O] = 0.4-0.77 (their HR numbers are indicated in Fig. 8), the model with $\Omega/\Omega_{crit} = 0.5$ and 0.7 ($V_0 \approx 220 - 300$ km/s) fits well. These results agree well with the data in section 3 on the rotation velocities of early B-type MS stars, specifically, about 80% of these stars have velocities of 0-130 km/s and only 20% have velocities of 200-400 km/s (18% in Fig. 8).

This study of the ratio N/O in early B-type MS stars shows that for the elements C, N, and O, at least in this specific case, the observations and theoretical predictions are in agreement. Another example of agreement between obser-vations and theory regarding the abundances of C, N, and O will be shown in the next section.

**6.3. The "nitrogen-carbon" anticorrelation for AFG-supergiants.** As noted above, after completion of the MS stage, early B-type stars rapidly enter the A-, F-, G-, and K-supergiant and giant stage. When a temperature $T_{eff} \leq$ 5900 K is reached, deep convective mixing (DCM) sets in and leads to further enhancement of the C, N, and O anomalies at the surface which appeared during the MS stage, in particular, the deficit of C and the excess of N. Note that the abundances of C, N. and O are determined from the C I, N I, and O I lines in these relatively cold stars.

The "nitrogen-carbon" anticorrelation for A-, F-, and G-supergiants has been known for over 30 years [43]. This anticorrelation was qualitatively understandable: during the CNO-cycle $^{12}C$ nuclei are converted into $^{14}N$ in a star's interior, so the carbon deficit should be accompanied by a nitrogen excess. A quantitative interpretation of this dependence was not, however, possible until the theoreticians moved from stellar models without rotation to models of rotating stars.

In Fig. 9 from [44], the observed "nitrogen-carbon" anticorrelation for the AFG-supergiants is compared with calculated [28] surface abundances of N and C for a model with mass $M = 12\,M_\odot$ and different initial rotation velocities $V_0$ (note that the relationship between N and C in these models depends weakly on the mass $M$). In the top frame the calculations correspond to the end of the MS stage and in the bottom, to the end of DCM in the AFG-supergiant stage. The values of $V_0$ are indicated next to the points.

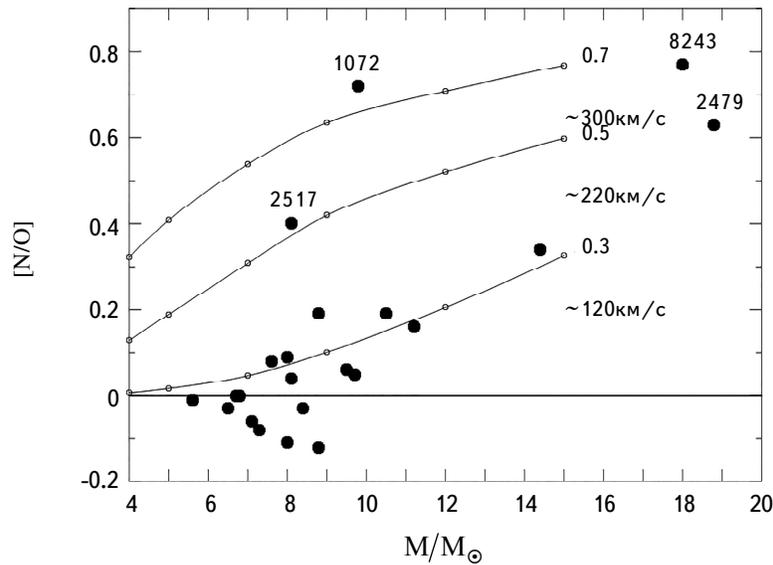

Fig. 8. [*N/O*] as a function of mass *M* for B-type stars at the end of the MS stage ($t/t_{MS}$ = 0.70-1.02) [42]. The smooth curves show theoretical dependences based on model calculations [14] for three values of the relative angular velocity $\Omega/\Omega_{crit} = 0.3$, 0.5, and 0.7.

An important conclusion follows from Fig. 9: *the 'nitrogen-carbon" anticorrelation mainly reflects the dependence of the anomalies in N and C on the initial rotation velocity $V_0$.* This fact could be established only with the aid of models of rotating stars.

In Fig. 9 the clustering of points within the marked grey square is noteworthy. Two possible explanations for this "cloud" are possible: (1) these are stars after the MS with initial velocities $V_0$ ~ 200-250 km/s (top frame) or (2) they are supergiants and giants close to finishing DCM with $V_0$ ~ 0-150 km/s (bottom frame). On recalling that 80% of stars with these masses ($M \approx 5-20 M_\odot$) had rotation velocities <150 km/s at the beginning of their evolution (see section 3), it becomes clear that the overwhelming majority of the stars in this cloud are supergiants and giants at the end of DCM.

Figure 9 (along with Fig. 8 in the previous section) illustrates the undoubted success of theory and explanations for the observed abundances of C, N, and O. In this regard, it should be noted that over the years the calculations with rotating star models have primarily been aimed, in particular, at explaining observations of C, N, and O. As the previous sections of this review show, this sort of agreement between theory and observations is often lacking. In particular, as pointed out above, the same model calculations which have been successful for C, N, and O, cannot explain the observed enhancement in the helium abundance in early B-type MS stars.

## 7. Conclusion

This review examines the first eight elements of the periodic table, H, He, Li, Be, B, C, N, and O. It is noteworthy that all these elements were originally synthesized in the Big Bang. The primordial abundances of the isotopes of these elements calculated using the standard model of the Big Bang (SMBB) are given here. For the primordial abundances of the isotopes of hydrogen and helium (deuterium, $^3$He, and $^4$He) there is excellent agreement between the SMBB and observational data, but in the case of lithium ($^7$Li) there is a discrepancy of 0.5 dex between the theory and observations. This follows from observations of old stars in the galactic halo, which have two features that have not yet been explained in the theory: the "lithium plateau" with a lithium abundance of $\log\varepsilon(\text{Li}) = 2.2$ (this value is 0.5 dex less than the predictions of the SMBB) and a significant reduction in this value for a number of stars.

The abundance of the light elements in stellar atmospheres depends on the initial rotation velocity, so typical rotation velocities of young stars in the MS stage have been examined. It is pointed out that, as opposed to traditional concepts, current data for hot stars (classes B and O) indicate a large number (about 80%) of stars with low rotation velocities $\leq 150$ km/s; this fact plays an important role in the interpretation of observed abundances of the light elements.

Given the large amount of data on the abundances of light elements in stars, primary emphasis has been on several unsolved problems, including the following:

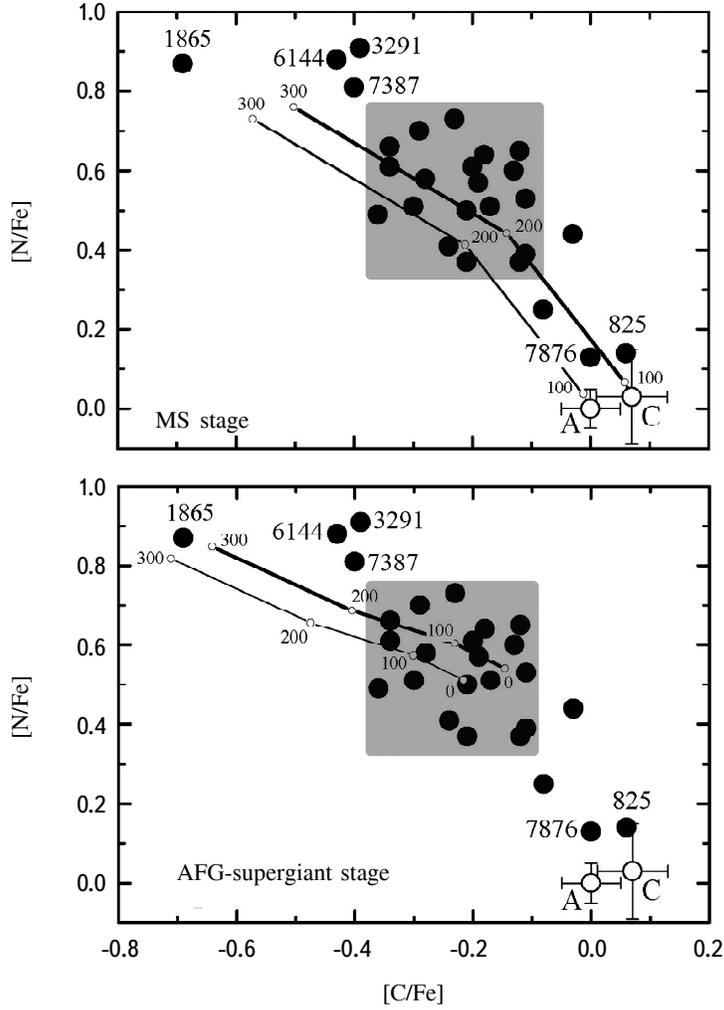

Fig. 9. The observed "nitrogen-carbon" dependence for AFG-supergiants compared with calculations for a model with $M = 12\,M_\odot$ and initial rotation velocities from 0 to 300 km/s [44]. The points A and C are the solar abundances of N and C [1,29].

**Helium.** The helium abundance He/H in early B-type MS stars manifests an increase with age; for the most massive stars with $M$ =12– 19 $M_\odot$, He/H rises toward the end of the MS by more than a factor of two. Theoretical models of stars in which rotationally induced mixing is taken into account cannot explain such a large increase in He/H. The sudden increase in He/H in the middle of the MS stage for B- and O-type stars which are components of close binary systems is a mystery for the theory.

Attempts have been made to explain the anomalous abundance of helium (as well as lithium) in the atmospheres of chemically peculiar MS stars, including magnetic stars (types He-s, He-w, HgMn, Am, and magnetic Ap), in terms of diffusion of atoms in surface layers. It is emphasized that the diffusion hypothesis must also simultaneously explain other features of the chemical composition of these stars, specifically a general trend toward excesses of elements with increasing atomic number $Z$, the distribution of spots of different elements on the surface of a star, and the distribution of different elements with depth.

**Lithium, beryllium, and boron.** These elements are burnt up in one and the same reaction (p, a), but at different temperatures (2.5, 3.5, and $5\times10^6$ K, respectively); thus, they show up in utterly different ways as indicators of stellar evolution. This is confirmed by observations and theoretically. For the FGK-dwarfs, the trend in the abundances of Li, Be, and B with falling $T_{eff}$, as well as the dip in the distribution of the abundances of these elements at $T_{eff}$ ~ 6600 K in the Hyades and other old clusters, have not been explained. Lithium rich FGK-giants and supergiants (with $\log\varepsilon(\text{Li})\geq 2$) are of special interest. Most of these have not been explained in terms of the standard theory of stellar evolution, so nonstandard hypotheses are of interest: recent synthesis of lithium in a star and capture by a star of a giant planet with the mass of Jupiter or greater.

**Carbon, nitrogen, and oxygen.** It has been found that superionization of C II, N II, and O II ions which has been neglected in the standard atmospheric models occurs in early B-type MS stars and late O-type stars. On the other hand, good agreement has been obtained for early B-type MS stars between the observed N/O ratios and calculations using rotating star models. In full agreement with the observed rotation velocity of these young stars, the value of N/O toward the end of the MS stage manifests an increase by 0.4-0.8 dex only for a small fraction (~20%) of the stars with sufficiently high initial rotation velocities of 200-400 km/s.

A quantitative explanation has been found for the well-known "nitrogen-carbon" anticorrelation in AFG-giants and supergiants: it reflects the dependence of the N and C anomalies on the initial rotation velocity $V_0$. It is noted that the same rotating star model calculations which have been successful in the case of C, N. and O cannot explain the observed increase in the abundance of helium in early B-type MS stars.